\documentclass[aps,physrev, preprint,preprintnumbers,longbibliography,a4paper,12pt,titlepage, nofootinbib,superscriptaddress]{revtex4-2}
\usepackage[utf8]{inputenc}
\usepackage[version=3]{mhchem}
\usepackage{braket}
\usepackage{amsmath,amssymb}
\usepackage{MnSymbol}
\usepackage{mathtools}
\usepackage{physics}
\usepackage{tensor}
\usepackage{empheq}
\usepackage{eufrak}
\usepackage{mathrsfs}
\usepackage{collectbox}
\usepackage{csquotes}
\usepackage{graphicx}% Include figure files
\usepackage{dcolumn}
\usepackage{bm}
\usepackage[colorlinks,breaklinks=true]{hyperref}
\hypersetup{
     colorlinks=true,
     linkcolor=blue,
     filecolor=blue,
     citecolor = black,      
     urlcolor=blue,
     }

\begin{document}

\preprint{}

\title{The Sugawara-Sommerfield construction including canonical spin current} % Force line breaks with \\
\author{Sandeep S. Cranganore}
\email{sandeep.cranganore@tuwien.ac.at}
\affiliation{Institute for Theoretical Physics, University of Cologne, 50923 Cologne, Germany}
\affiliation{TU Wien - Atominstitut,
Stadionallee 2, 1020 Vienna, Austria}%Lines break automatically or can be forced with \\
\date{\today}% It is always \today, today,
\begin{abstract}
We set up the Sugawara-Sommerfield (SS) construction and generalize it by the inclusion of canonical spin current. Using the techniques of current algebra, we infer that the canonical spin current are linear in the vector-axial vector currents. From a geometric perspective, the underlying manifold has a vanishing Lorentz curvature and a non-vanishing torsion. This leads to teleparallelism and the canonical spin current (connection) assume a pure gauge form. Moreover, this model provides a possibility to unify gravity with strong-interactions by expressing the gravitational gauge connections in terms of the Yang-Mills gauge connections.
\end{abstract}

\maketitle

%\tableofcontents

\section*{NOTATIONS}
Spacetime coordinates will be labeled with Latin indices $i, j, k...=0, 1, 2, 3$. Spatial coordinates will be denoted by a, b,....= 1, 2, 3. The frame (tetrad) fields $e_\alpha$ with $\alpha$, $\beta,...=0,1,2,3$ (Lorentz indices)  with components $\tensor{e}{^i_\alpha}$,   coframe field  $\vartheta^\beta$ with components (vierbeins) $\tensor{e}{_j^\beta}$. $\omega_\alpha =\ast \vartheta_\alpha$, $\tensor{\omega}{_\alpha_\beta} = \ast (\vartheta_\alpha \wedge \vartheta_\beta)$, $\tensor{\omega}{_\alpha_\beta_\gamma} = \ast (\vartheta_\alpha \wedge \vartheta_\beta \wedge \vartheta_\gamma)$,  $e:= det (\tensor{e}{_j^\beta}) = \sqrt{-g}$, $c=1$.  Parentheses around the indices denote symmetrization $(ij):=\frac{1}{2}(ij+ji)$ and antisymmetrization $[ij]:=\frac{1}{2}(ij-ji)$. The covariant exterior derivative components are $D_\alpha=\tensor{e}{^i_\alpha}D_i$ with $D_i=\partial_i+\tensor{\Gamma}{_i^\alpha^\beta}\tensor{\Sigma}{_\alpha_\beta}$, where $\tensor{\Gamma}{_i^\alpha^\beta}$ is the Lorentz (spin) connection and $\tensor{\Sigma}{_\alpha_\beta}$ are the representations of the Lorentz generators. The metric field will be denoted as $g_{ij}(x)$ and the Minkowski metric as $\eta_{ij}$, while the three-metric reads $h_{ab}(x)$. Internal symmetry (flavour) indices are labeled as A, B, C,...,. $T^A$ are the  generators of internal symmetry groups (for e.g., $SU(2)_f$ (Isospin) or $SU(3)_f$ flavour groups). $\epsilon_{ABC}$ are the completely antisymmetric Levi-Civita symbol. $A = \tensor{A}{^A_i}(x) T_A  dx^i$ are the Yang-Mills gauge connection $1$-forms. Coframe field with \textit{internal symmetry indices} is denoted as $\vartheta^A(x)$ and their components as  $\tensor{E}{_i^A}(x)$. ($\tensor{J}{^A_i}\rightarrow \tensor{J}{_5^A_i}$) corresponds to similarly for the axial currents. $\widetilde{J}$ corresponds to either vector or axial vector currents. 
\cleardoublepage

\section{Introduction}
In the year 1968, both H.Sugawara \cite{PhysRev.170.1659} and C.M. Sommerfield \citep{PhysRev.176.2019} independently developed a theory which was purely based on currents as dynamical quantities along with its associated commutation rules. This model is equivalent to a \enquote{formal limit} of a massive Yang-Mills theory, where the bare mass $m$ and bare coupling $g$ both tend to zero and their ratio to a constant \cite{PhysRev.170.1353}. The commutation rules come from \enquote{Current Algebra}\footnote{In mathematics, especially for 2-dimensional systems, current algebra led to the birth of affine Kac-Moody algebra. Hence we shall sometimes use this term interchangeably in our paper.} which are nothing but equal time commutation relations (ETCRs) between the currents. This was first put forth by M. Gell-Mann in $1962$, in order to describe the strong interaction between hadrons \cite{cao_2010} and was later developed into a complete set of commutation rules in \cite{PhysRevLett.18.1029} using a massive Yang-Mills theory. The complete set of hadronic currents are the \textit{vector} currents $\tensor{J}{^A_i}(x)$ and the \textit{axial vector} currents $\tensor{J}{_5^A_i}(x)$, entailing the approximate $SU(3)_f$ flavour symmetry of strong interactions. These (Lie-algebra valued) $\mathfrak{su}(3)$ - currents are closed under ETCRs and satisfy the $SU(3)\otimes SU(3)$ algebra. A very important result of the Sugawara-Sommerfield (SS) construction was that the \enquote{\textit{Hilbert symmetric energy-momentum (current) tensor} $\Theta_{ij}(x)$ which is the current coupled to the metric field $\tensor{g}{_i_j}(x)$ turned out to be a \textit{bilinear} expression in $J_i(x),J_{5i}(x)$}. This was the first time that the energy-momentum tensor featuring in Einstein's General Relativity (GR) theory, which is based on external spacetime symmetries (for e.g., diffeomorphism invariance \cite{kiefer2012quantum}, Lorentz invariance), could be solely expressed in terms hadronic currents based on internal flavour and gauge symmetries. %Hadron physics extensively stresses the importance of currents associated with internal gauge symmetries.
%The central idea behind this scheme was that currents are the measurable quantities entailing the approximate $SU(3)_f$ flavour symmetry of strong interactions and was proposed in the eightfold way.  %According to the Noether theorem, symmetries imply conservation laws,  but this approach suffered drawbacks since the flavour symmetry is an approximate symmetry and works satisfactorily only for the light quarks $(u,d,s)$.  Another way to describe the underlying symmetry was in terms of a Lie algebra. The complete set of the hadronic currents are the \textit{vector} currents $\tensor{J}{^A_i}(x)$ and the \textit{axial vector} currents $\tensor{J}{_5^A_i}(x)$. These $\mathfrak{su}(3)$ - currents are closed under the equal time commutation relations (ETCRs) and satisfy the $SU(3)\otimes SU(3)$ algebra. In conformal field theories these relations are known as the affine Kac-Moody algebra.
\par

It is very well known that GR is far from explaining the nature of gravitational fields in the quantum mechanical regime. One such viable theory and a more complete description of gravity is the Einstein-Cartan (EC) theory. In EC gravity it is not only the mass but also the spin of a quantum particle that distorts spacetime. It takes place in a manifold called the Riemann-Cartan (RC) spacetime. This theory ideally serves as a precursor to a quantum theory of gravity since it is built on the idea of the Wigner classification in quantum mechanics. The Wigner's classification states that quantum particles are irreducible unitary representations of the Poincaré group and are uniquely labeled by their mass (inertia) and their intrinsic angular momentum (spin) which are related to the two invariants of the Lie algebra of the Poincaré group.  Mass is linked to the 4-dimensional translational group  $T_4$ and the spin is linked to the 6-dimensional Lorentz group $SO(1,3)$. 
\newline The Poincaré group $ P(1,3) = T_4\rtimes SO(1,3)$ is their semi-direct product and is a 10-dimensional non-compact Lie group \cite{doi:10.1142/p781}. The EC theory is obtained by gauging the Poincaré group, hence, it is a gauge theory of gravity. The sources of gravity (geometry) in the EC theory are the canonical Noether currents namely the canonical energy-momentum tensor (current) linked to the translational group and the canonical spin tensor (current) linked to the Lorentz group. The canonical energy-momentum current $\mathfrak{T}_\alpha (x)$ are coupled to the translational gauge connection (coframe field) $\vartheta^\alpha (x)$ and are the sources of curvature while the canonical spin current $\tensor{\mathfrak{S}}{_\alpha_\beta}(x)$ are coupled to the spin (Lorentz) connection $\tensor{\Gamma}{^\alpha^\beta} (x)$ and are the sources of non-propagating torsion. For a particle of mass $m$ and a reduced Compton wavelength $\lambda_{Co} := \frac{\hbar}{mc}$, the effects of torsion becomes dominant at the critical length, $r_{EC} \sim (\lambda_{Co} l_P^2)^{\frac{1}{3}}$ ($l_P =$ Planck length) and also in the very early Universe  \citep{kiefer2012quantum, doi:10.1142/p781}. Hence these two currents are the sources of the gravitational field in the microphysical realm. For a consistent quantization of the gravitational field, a gauge description is often overlooked. Moreover, fermionic matter being spin-$\frac{1}{2}$ fields, massive spin-1 gauge bosons like the $W^{\pm}, Z^0$ have non-vanishing spin tensor and has a feature of spin-spin contact interaction. In fact, the tensor, vector and axial vector parts of the spin current act as sources of massive spin-$2$ and spin-$1$ massive neutral mesons respectively along with the helicity-$2$ gravitons sourced by the metric field. Hence, they also provide viable candidates for an extended standard model. Due to the aforementioned reasons it is of primary importance to comprehensively study the gravitational currents/sources - ($\mathfrak{T},\mathfrak{S}$) which encode the underlying Poincaré gauge symmetry. Hence, a complete description of a theory of hadrons actually requires \big($J_i(x),J_{5i}(x), \tensor{\mathfrak{T}}{_i^j}(x), \tensor{\mathfrak{S}}{_i_j^k}(x)\big)$.
\par
%These th of vector currents are (i) the isoscalar electromagnetic current, (ii) the triplet of isovector electromagnetic currents, (iii) the weak strangeness conserving currents and (iv) the two doublets of weak strangeness changing currents.  corresponds to an $SU(3)\otimes SU(3)$ algebra. The set of leptonic currents also satisfy a similar algebra. A mathematical generalization of the $SU(3)$ current algebra is the affine Kac-Moody algebra \cite{PMIHES_1998__S88__153_0}. The corresponding currents associated to the internal $SU(3)_f$ transformations are the octet of vector and axial vector currents.
Motivated by the fact that the currents of EC theory are the two actual sources of gravity, based on (i) a gauge principle and (ii) closer to a quantum description of gravity, we should seriously consider extending the SS construction for these two currents. In the XV Advanced Research Workshop on High Energy Spin Physics, Dubna, 2013 \cite{hehl2014energymomentum}, F. W. Hehl posed the following problem \enquote{Schwinger (1963) studied, for example, the equal time commutators of 
the components of the Hilbert energy-momentum tensor [36]. Should one 
try to include also the spin tensor components and turn to the canonical tensors? In the Sugawara model (1968), \lq A field theory of currents\rq\  was 
proposed with 8 vector and 8 axial vector currents for strong 
interaction and a symmetric energy-momentum current for gravity that 
was expressed bilinearly in terms of the axial and the vector 
currents. Now, when we have good arguments that the gravitational 
currents are $\mathfrak{T}_\alpha$ and $\mathfrak{S}_{\alpha\beta}$, one may want to develop a corresponding current algebra by determining the equal time commutator of these 
currents....}
\subsection{Plan of the paper}
In this paper, we first derive the current algebra (Kac-Moody algebra) for the gravitational currents, namely the canonical spin current and canonical energy-momentum current. Later, using the set of current commutation rules in \cite{PhysRev.170.1659}, we make the SS ansatz for the canonical spin current. It follows from the Belinfante-Rosenfeld symmetrization that the canonical spin current is a linear polynomial in the hadronic vector and axial vector flavour currents. Since a current-gauge potential relation exists in this model, we find an expression for the affine spin connection in terms of the massive Yang-Mills gauge connection.  This further deepens the analogy between spin and charge or flavour, which are borne out by gauge principles. 
This leads to an inference that, the SS construction provides a method for unifying gravitational gauge theories and the strong and weak interactions. Thus, it is possible to represent the sources or gauge potentials associated to external spacetime symmetry by sources or gauge potentials associated to internal symmetries. We later try to cast the spin current ansatz as a unitary singlet in terms of the internal-vierbeins $\tensor{E}{_i^A}(x)$ and set up its canonical quantization.
\par
Interestingly, we find that the SS construction for spin current leads to non-intuitive implications to the underlying geometry of the manifold, namely \enquote{teleparallelism}. Teleparallelism or absolute parallelism is a feature of a manifold with vanishing Lorentz curvature but non-vanishing torsion. The underlying arena is called a \enquote{Weitzenböck spacetime}.  We thrown some light on its surprising emergence in our case.
\par
Then, we discuss a very interesting analogy which exists between our extended model and the non-linear $\sigma$ model with Wess-Zumino-Witten (WZW) interactions. we make a comparison between our 4-dimensional spacetime construction and this 2-dimensional theory, where it was shown that the renormalization group structure of models with Wess-Zumino terms allowed for an elegant geometrical interpretation by incorporating torsion into the manifold \cite{PhysRevLett.53.1799}. It was further shown in \cite{BRAATEN1985630} that by including the Wess-Zumino interactions in the non-linear $\sigma$ model, the renormalization flow leads to a deformation of the geometry of the field manifold, wherein the rate of change of metric is also proportional to the asymmetric Ricci tensor which contains contributions from the torsion tensor. The manifold was shown to be \enquote{parallelizable} (vanishing Lorentz curvature and non-vanishing torsion) at special values of the coupling $\lambda = \pm 4\pi /N$ for $N \in \mathbb{Z}$. These are the infrared fixed points of the theory. In the pure bosonic model these coincide with the zeros of the $\beta$ function. The non-trivial infrared fixed points occur as a consequence of teleparallelism for the renormalized geometry and is known as \enquote{Geometrostasis}.
\par
Lastly, we discuss very briefly, where the SS construction for spin can be applied and similar features that exists between this model and in condensed matter systems.
\section{CURRENT COMMUTATORS FOR CONSTRUCTING THE MODEL}\label{CA}
\subsection{The complete set of current commutators}
In order to make a construction à la Sugawara-Sommerfield for the spin current, we need a set of commutators for the $\mathfrak{su}(2)$-valued $4$-currents (hadronic currents).
The time-time ETCRs,
\begin{subequations}\label{paper01_eq:1}
\begin{equation}
 [\tensor{J}{^A_0}(x),\tensor{J}{^B_0}(x')]_{x_0=x'_0}=i\tensor{\epsilon}{_A_B_C} \tensor{J}{^C_0}(x)\delta^3(x-x'),    
\end{equation}
\begin{equation}
[\tensor{J}{^A_0}(x),\tensor{J}{_5^B_0}(x')]_{x_0=x'_0}=i\tensor{\epsilon}{_A_B_C }\tensor{J}{_5^C_0}(x)\delta^3(x-x'),
\end{equation}
\begin{equation}
[\tensor{J}{_5^A_0}(x),\tensor{J}{_5^B_0}(x')]_{x_0=x'_0}=i\tensor{\epsilon}{_A_B_C}\tensor{J}{^C_0}(x)\delta^3(x-x').
\end{equation}
\end{subequations}
\newline
Note that we shall only concentrate on $\mathfrak{su}(2)$ and not the current octet related to $\mathfrak{su}(3)$.
\newline
The time-space ETCRs for the currents were set up from the Algebra of fields \cite{PhysRevLett.18.1029},
\begin{subequations}\label{paper01_eq:2}
\begin{equation}
[\tensor{J}{^A_0}(x),\tensor{J}{^B_a}(x')]_{x_0=x'_0}= i\tensor{\epsilon}{_A_B_C } \tensor{J}{^C_a}(x)\delta^3(x-x')+iC\tensor{\delta}{^A^B}\partial_a \delta^3(x-x'),
\end{equation}
\begin{equation}
[\tensor{J}{_5^A_0}(x),\tensor{J}{_5^B_a}(x')]_{x_0=x'_0}= i\tensor{\epsilon}{_A_B_C }\tensor{J}{^C_a}(x)\delta^3(x-x')+iC\tensor{\delta}{^A^B}\partial_a\delta^3(x-x'),
\end{equation}
\begin{equation}
[\tensor{J}{^A_0}(x),\tensor{J}{_5^B_a}(x')]_{x_0=x'_0}= i\tensor{\epsilon}{_A_B_C}\tensor{J}{_5^C_a}(x)\delta^3(x-x'),    
\end{equation}
\begin{equation}
[\tensor{J}{_5^A_0}(x),\tensor{J}{^B_a}(x')]_{x_0=x'_0}= i\tensor{\epsilon}{_A_B_C} \tensor{J}{_5^C_a}(x)\delta^3(x-x').    
\end{equation}
\end{subequations}
\newline
Here $A,B,C = 1,2,3$, $C=\frac{m^2}{g^2}$ is a c-number with length dimension $d_C=-2$ \cite{PhysRevLett.18.1029}. The space-space commutators are all identically equal to zero. The derivatives of the delta function correspond to the Schwinger terms which are proportional to $\hbar^2$. Mathematically, this corresponds to an  $SU(2)\otimes SU(2)$ current algebra.
\newpage
\subsection{The Sugawara-Sommerfield model}\label{section 1} \label{SR}
Originally, in the Sugawara model the octet of vector currents $\tensor{J}{^A_i}(x)$ and the octet of axial vector currents $\tensor*{J}{_5^A_i}(x)$ entail the underlying symmetry and play the role of dynamical variables. In our further discussions we will work with current triplet and not octet.  This means that the currents $(J^A, \tensor{J}{_5^A})$ play the role of coordinates labelling the hadrons. This model is a \enquote{formal} limit of a massive Yang-Mills theory described by the Lagrangian density \cite{PhysRev.170.1353},
\begin{align}\label{proca}
\mathfrak{L}(A,\partial A, x)=-\frac{1}{4}\tensor{F}{^A_i_j}(x)\tensor{F}{^A^i^j}(x)+\frac{m^2}{2}\tensor{A}{^A_i}(x)\tensor{A}{^A^i}(x),
\end{align}
where the components of the field strength $2$-forms are,
\begin{align*}
\tensor{F}{_i_j^A}(x)= 2\partial_{[i} \tensor{A}{_j_]^A}(x) -g\tensor{\epsilon}{_A_B_C}\{\tensor{A}{_i^B}(x),\tensor{A}{_j^C}(x)\}.    
\end{align*}
\par
Here, $\{,\}$ is the anticommutator. We define the currents $J$ (also applicable to $J_5$) by performing a scale transformation on the spin-$1$ massive gauge fields $\tensor{A}{_i^A}(x)$ \cite{PhysRev.170.1353},
\begin{equation*}
\tensor{A}{_i^A}(x) = \frac{g}{m^2} \tensor{J}{^A _i}(x).
\end{equation*}
\begin{equation*}
\tensor{\widetilde{F}}{_i_j^A}(x)=\frac{g}{m^2}\tensor{F}{_i_j^A}(x).
\end{equation*}
Letting the bare mass $m\rightarrow 0$, bare coupling $g\rightarrow 0$ and their ratio $\frac{m^2}{g^2}\rightarrow C$, the canonically conjugate momenta become, 
\begin{equation*}
\lim_{m\rightarrow 0, g\rightarrow 0,\frac{m^2}{g^2}=C} \tensor{\widetilde{F}}{_0_i^A}(x) = 0.
\end{equation*} 
This means $\widetilde{F}$ vanishes, like $m^2$, in this limit. Hence, this doesn't reduce it to the vector meson theory \cite{PhysRev.170.1659},

By the same argument, the spatial components of the field strengths $\tensor{\widetilde{F}}{_a_b^A}(x)$ also vanish in the prescribed limit. Thus, the canonically conjugate momenta completely vanish, leaving us with only the currents. This boils down to a \textit{noncanoncial} formalism. Let us stress that the conjugate momenta vanish only in this limit and exist for all other values of $m$,$g$.
\subsection{The SS construction for the symmetric energy-momentum current revisited}
One can construct the symmetric (Hilbert) energy-momentum current $\Theta(x)$ in terms of the hadronic currents by using the following conditions:
\newline
(i) The ETCRs (Schwinger commutators) for the energy density, which fixes the polynomial order of the currents \cite{PhysRev.130.406}, 
\begin{equation}\label{Schw}
[\tensor{\Theta}{^0^0}(x),\tensor{\Theta}{^0^0}(x')]= -i \big(\tensor{\Theta}{^0^a}(x) + \tensor{\Theta}{^0^a}(x')\big)\partial_a \delta^3(x-x').
\end{equation}
(ii) $\Theta(x)$ should be a unitary singlet, i.e.\ it should be invariant under the internal (approximate) $SU(3)_f$, $SU(2)_f$ symmetry transformations.
\newline
(iii) The conservation of the energy-momentum tensor, $\partial_i\tensor{\Theta}{^i^j}(x)=0$.
\newline
(iv) Poincaré invariance.
\newline

It follows from the above conditions and Eqs.\  (\ref{paper01_eq:1}, \ref{paper01_eq:2}) that the symmetric energy-momentum current are restricted to be bilinear in the hadronic currents,
\begin{equation} \label{paper01_eq:3} 
\tensor{\Theta}{_i_j}(x) = -\frac{1}{2C}[ \{\tensor{J}{^A_i}(x),\tensor{J}{^A_j}(x)\}- \tensor{g} {_i_j}(\tensor{J}{^A_k}(x) \tensor{J}{^A^k}(x))+ \big(\tensor{J}{^A_i}\rightarrow \tensor{J}{_5^A_i}\big)].
\end{equation}

This is an interesting construction for the symmetric energy-momentum current for the theory of strong interactions as it interlocks internal symmetry $SU(3)\otimes SU(3)$ Lie-algebra, Poincaré invariance and Schwinger terms \cite{PhysRev.170.1353}. In the case of spinor fields, the currents $\tensor{J}{_i}(x)=\bar{\psi(x)}\gamma_i\psi(x)$ are quadratic in the fields $\psi (x)$ and $\bar{\psi}(x)$. This results in the energy-momentum tensor being \textit{quartic} in the fields, which suffers severly from short distance singularities \cite{PhysRev.180.1359}.
\par 
One can also arrive at Eq.\  (\ref{paper01_eq:3}) and the equations of motion described below Eqs.\  (\ref{paper01_eq:4}, \ref{paper01_eq:5}) by setting up the Yang-Mills equations of motion and the symmetric energy momentum tensor using Eq.\  (\ref{proca}) and then applying the limit \cite{PhysRev.170.1353}.
\subsection{Heisenberg equations of motion for the currents}
For the theory to be quantum mechanical, the currents should satisfy the Heisenberg equations of motion \cite{PhysRev.170.1659},
\begin{equation*}
[p_i,\tensor{\widetilde{J}}{^A_j}(x)] = -i\partial_i\tensor{\widetilde{J}}{^A_j}(x). 
\end{equation*}
Here, $p_i = \int d^3x \tensor{\Theta}{_0_i}(x)$ is the $4$-momentum.

The currents satisfy the \textit{Lorenz} gauge condition (current conservation),
\begin{align}\label{paper01_eq:4}
\partial_i\tensor{J}{^A_i}(x) = 0 && \partial_i\tensor{J}{_5^A_i}(x) = 0.
\end{align}
As a result of a vanishing field strength, as mentioned in \cite{PhysRev.170.1353}, the derivatives boil down to algebraic (bilinear) expressions \cite{PhysRev.170.1659},
\begin{subequations}\label{paper01_eq:5}
\begin{equation}\label{paper01_eq:5a}
\partial_i\tensor{J}{^A_j}(x)-\partial_j\tensor{J}{^A_i}(x)= \frac{1}{2C}\tensor{\epsilon}{_A_B_C}\big(\tensor{J}{^B_i}(x)\tensor{J}{^C_j}(x)+\tensor{J}{^C_j}(x)\tensor{J}{^B_i}(x)\big)+ (\tensor*{J}{^A_i}\rightarrow \tensor*{J}{_5^A_i}),
\end{equation}
\begin{equation}\label{paper01_eq:5b}
\partial_i\tensor{J}{_5^A_j}(x)-\partial_j\tensor{J}{_5^A_i} (x) =  \frac{1}{C}\tensor{\epsilon}{_A_B_C}\big( \{ \tensor{J}{^B_i}(x), \tensor{J}{_5^C_j}(x)\}+ \{\tensor{J}{_5^B_i}(x), \tensor{J}{^C_j}(x)\}\big). 
\end{equation}
\end{subequations}
If the axial vectors don't play a role in the theory,  Eq.\  (\ref{paper01_eq:5a}) can be expressed as a covariant derivative,  with the currents themselves acting as the gauge connection (potential),
\begin{equation*}
\tensor{F}{_i_j^A} (x) = \tensor{D}{_i^A^B}\tensor{J}{^B_j}(x)-\tensor{D}{_j^A^B}\tensor{J}{^B_i}(x) = 0.
\end{equation*}
Here $\tensor{D}{_i^A^B} = \tensor{\delta}{^A^B}\partial_i-\frac{1}{2C}\tensor{\epsilon}{_A_C_B}\tensor{J}{^C_i}(x)$ is the Yang-Mills like covariant derivative.
The field equation of the Sugawara model is,
\begin{equation*}
\partial_i\tensor{F}{^i^j^A}(x)= 0. 
\end{equation*}
This model applies to $SU(2)$, $SU(3)$, chiral $SU(2)_L \otimes SU(2)_R$, chiral $SU(3)_L \otimes SU(3)_R$ (i.e. only to \textit{compact} Lie groups) \cite{PhysRev.180.1359}.
\par
Corresponding to every Sugawara model is an associated Lagrangian field theory (Bardacki-Halpern construction) and vice-versa \cite{PhysRev.180.1359}. The action is quadratic in the currents. 
\begin{equation}\label{paper01_eq:6}
S = (1/2C) \int d^4x (\sqrt{-g})\tensor{g}{^i^j} \tensor{J}{^A_i}\tensor{J}{^A_j} + (\tensor{J}{^A_i}\rightarrow \tensor{J}{_5^A_i}).
\end{equation}
The Hilbert definition of the symmetric energy-momentum current, i.e., $\tensor{\Theta}{_i_j}(x) := -\frac{2}{\sqrt{-g}} \frac{\delta S}{\delta \tensor{g}{^i^j}(x)}$, yields Eq.\  (\ref{paper01_eq:3}). 
\subsection{Schwinger commutators revisited}
It was shown by Schwinger that the symmetric energy-momentum components obey a certain set of local equal-time commutation relations. We shall record some of the commutation relations for the symmetric energy-momentum tensor \citep{PhysRev.130.406, schwinger1970particles},
\begin{subequations}\label{paper01_eq:7}
\begin{equation}
[\tensor{\Theta}{^0^0}(x),\tensor{\Theta}{^0^0}(x')]= -i \big(\tensor{\Theta}{^0^a}(x)+\tensor{\Theta}{^0^a}(x')\big)\partial_a \delta^3(x-x') 
-\tensor{\bar{\tau}}{^0^0^,^0^0}(x,x'),
\end{equation}
\begin{equation}
[\tensor{\Theta}{^0^0}(x),\tensor{\Theta}{^0^a}(x')]= -i \big(\tensor{\Theta}{^a^b}(x) +\tensor{\Theta}{^0^0}(x')\tensor{\delta}{^a^b}\big)\partial_b \delta^3(x-x')-\tensor{\bar{\tau}}{^0^0^,^0^a}(x,x'),   \end{equation}
\begin{equation}
[\tensor{\Theta}{^0^a}(x),\tensor{\Theta}{^0^b}(x')] = -i\big(\tensor{\Theta}{^0^b}(x)\partial^a  +\tensor{\Theta}{^0^a}(x')\partial^b\big)\delta^3(x-x')  -\tensor{\bar{\tau}}{^0^a^,^0^b}(x,x'). \end{equation}
\end{subequations}
Here, the operators $\bar{\tau}$ are model dependent Schwinger terms and are constrained to have  certain integrals and moments to vanish \cite{doi:10.1063/1.1705368}, although, they may be non-vanishing for several cases. We first set up the Schwinger commutators for the canonical spin and energy-momentum current.
\section{{EINSTEIN-CARTAN (EC) THEORY, CURRENTS, NOETHER IDENTITIES AND SCIAMA-KIBBLE EQUATIONS}} \label{EC}
The introduction of a Riemann-Cartan geometry leads to a gauge theory of gravitation. The gauge potentials of the RC spacetime are the coframe fields  $\vartheta^\alpha=\tensor{e}{_i^\alpha}dx^i$, which are $4$ linearly independent 1-forms having ($4\cross4$) components, acting as the translational gauge potential. The Lorentz connection $\tensor{\Gamma}{^\alpha^\beta}=\tensor{\Gamma}{_i^\alpha^\beta} dx^i = -\tensor{\Gamma}{^\beta^\alpha}$ consists of $4$ linearly independent 1-forms having ($4\cross6$) components, acts as the rotational gauge potentials \cite{doi:10.1142/p781}.
\par
The translational field strength associated to $\vartheta^\alpha$ is the torsion $2$-form  ($4\cross 6 = 24$ components) defined as, 
\begin{align}\label{trans}
T^\alpha := D\vartheta^\alpha =
d\vartheta^\alpha+\tensor{\Gamma}{^\alpha_\beta}\wedge \vartheta^\beta=
\frac{1}{2} \tensor{T}{_i_j^\alpha} \vartheta^i\wedge\vartheta^j,
\end{align}
where D is the covariant exterior derivative with components obtained by taking interior product $e_\alpha \righthalfcup D = D_\alpha$.
The rotational field strength associated to $\tensor{\Gamma}{^\alpha^\beta}$ is the Lorentz curvature $2$-form (having $6\cross6=36$ components) defined as,
\begin{equation}\label{rots}
\tensor{R}{^\alpha^\beta}:= d\tensor{\Gamma}{^\alpha^\beta}-\tensor{\Gamma}{^\alpha^\gamma}\wedge\tensor{\Gamma}{_\gamma^\beta}=\frac{1}{2}\tensor{R}{_i_j^\alpha^\beta}\vartheta^i\wedge\vartheta^j.
\end{equation}
\par
The introduction of the gauge potentials and the corresponding field strengths leads to the generalization of the Lie-algebra of the Poincaré group \cite{Hehl:1994ue}, \begin{subequations}\label{paper01_eq:8}
\begin{equation}
[\tensor{\Sigma}{_\alpha_\beta} , D_\gamma]  =\tensor{g}{_\gamma_\alpha}D_\beta - \tensor{g}{_\gamma_\beta}D_\alpha ,
\end{equation}
\begin{equation}
[D_\alpha,D_\beta] =\frac{1}{2}\tensor{R}{_\alpha_\beta^\gamma^\delta}\tensor{\Sigma}{_\delta_\gamma}-\tensor{T}{_\alpha_\beta^\gamma}D_\gamma ,
\end{equation}
\begin{equation}
[\tensor{\Sigma}{_\alpha_\beta},\tensor{\Sigma}{_\gamma_\delta}] = i\big(\tensor{\eta}{_\beta_\gamma}\tensor{\Sigma}{_\alpha_\delta}-\tensor{\eta}{_\alpha_\gamma}\tensor{\Sigma}{_\beta_\delta}+\tensor{\eta}{_\alpha_\delta}\tensor{\Sigma}{_\beta_\gamma}-\tensor{\eta}{_\beta_\delta}\tensor{\Sigma}{_\alpha_\gamma}\big) .  
\end{equation}
\end{subequations}
 Here,
$\tensor{R}{_\alpha_\beta^\gamma^\delta} = \tensor{e}{^i_\alpha}\tensor{e}{^j_\beta}\tensor{R}{_i_j^\gamma^\delta}$ \quad\text{and}\quad $\tensor{T}{_\alpha_\beta^\gamma}=\tensor{e}{^i_\alpha}\tensor{e}{^j_\beta}\tensor{T}{_i _j^\gamma} $. \subsection{Noether currents (sources) of the EC theory}
Matter current $3$-forms are obtained via the Lagrange-Noether machinery by requiring invariance under local Poincaré transformations  (with the matter equation  $\frac{\delta\mathfrak{L}}{\delta{\psi}}=0$ satisfied) \cite{Hehl:1994ue}. The currents of EC theory coincide with the Noether currents associated to external local spacetime transformations \cite{Obukhov:2018bmf}. 
The Noether current associated to the translational gauge connection is the canonical energy-momentum current,
\begin{subequations}\label{paper01_eq:9}
\begin{equation}
\tensor{\mathfrak{T}}{_\alpha^i}:= -\frac{\delta \mathfrak{L}}{\delta\tensor{e}{_i^\alpha}} \equiv \frac{\partial\mathfrak L}{\partial(D_i\psi)}D_\alpha\psi - \tensor{e}{^i_\alpha}\mathfrak{L}.
\end{equation} 
The current associated to the rotational gauge connection are the canonical spin current,
\begin{equation}
\tensor{\mathfrak S}{_\alpha _\beta^i}:=-2\frac{\delta\mathfrak L}{\delta\tensor{\Gamma}{_i^\alpha^\beta}} \equiv  - \frac{\partial\mathfrak L}{\partial D_i\psi}\tensor{\Sigma}{_\alpha_\beta}\psi. 
\end{equation}
\end{subequations}
The components of the spin current are Lie-algebra (Lorentz)  valued  $3$-forms. 
\newline 

It is very important to note that for a consistent definition of global charges  (spin and charge/flavour charges), we require that the Noether currents satisfy an \textit{on-shell} conservation law. This means that are currents are localized and \textit{vanish} over the hypersurface bounding them. 
\subsubsection{The spin angular momentum algbera}
Like charge is the spatial integral over the time component of the 4-current density, spin angular momentum is the integral over the time component of the spin current density, i.e.,
\begin{equation}
    S_a := \frac{1}{2}\epsilon_{abc} \int   \tensor{\mathfrak{S}}{_b_c^0}(x) d^3x.
\end{equation}
It is easy to show that $S_a$ satisfy the spin angular momentum algebra of quantum mechanics, 
\begin{equation}
    [S_a,S_b] = i\epsilon_{abc} S_c.
\end{equation}
\subsection{Conservation laws and Noether identities}
The invariance of the Lagrangian under local diffeomorphisms yields us the \textit{first Noether identity} for the canonical energy-momentum current \cite{Hehl:1994ue},
\begin{equation}\label{paper01_eq:10}
D_i \tensor{\mathfrak{T}}{_\alpha^i} \equiv \tensor{T}{_\alpha_i^\beta}\tensor{\mathfrak{T}}{_\beta^i}+\frac{1}{2}\tensor{R}{_\alpha_i^\gamma^\delta}\tensor{\mathfrak S}{_\gamma_\delta ^i}.
\end{equation}
The above identity for the canonical energy-momentum implies that the field strengths act upon the corresponding sources.

Invariance under infinitesimal Lorentz transformations yields the Noether identity for the canonical spin current, 
\begin{equation} \label{paper01_eq:11}
D_i\tensor{\mathfrak S}{_\alpha _\beta^i} \equiv 2\tensor{\mathfrak{T}}{_{[\alpha \beta]}}.
\end{equation}

The above identities in the special relativistic setting are the standard conservation laws. The  Noether identities would be central in deriving the commutation relations for the canonical energy-momentum and the spin current in this paper.
\subsection{Field equations of EC theory}
The field equations of EC theory can be derived from an action principle. The Lagrangian consists of the field Lagrangian and the matter Lagrangian. The EC Lagrangian density is linear in the Lorentz curvature (invariant). The matter Lagrangian density is a scalar-valued $4$-form (top-form) \cite{Obukhov:2018bmf},
\begin{align}
S_{EC} = \int d^4x \bigg( \frac{1}{2\kappa}\tensor{\omega}{^\alpha^\beta} \wedge \tensor{R}{_\alpha_\beta} +  \mathfrak L_m(\psi,d\psi,\vartheta^\alpha, \tensor{\Gamma}{^\alpha^\beta}) \bigg) .
\end{align}
Variation w.r.t.\ $\Gamma^{\alpha\beta}$ yields an algebraic equation, with spin as the source of torsion,
\begin{equation}
\frac{1}{2}\tensor{\omega}{_{\alpha \beta \gamma}} \wedge T^\gamma = \kappa \tensor{\mathfrak{S}}{_\alpha_\beta}.
\end{equation}
Similarly, variation w.r.t.\ $\vartheta^\alpha$ yields the very familiar equation for curvature,  
\begin{equation}
\frac{1}{2}\tensor{\omega}{_{\alpha \beta \gamma}} \wedge \tensor{R}{^\beta^\gamma} = \kappa \mathfrak{T}_\alpha.  
\end{equation}
It is the \textit{canonical energy-momentum current} which acts as a source of curvature in EC theory, instead of the symmetric (Hilbert) energy-momentum current. \cleardoublepage
\section{CONSTRUCTION OF ETCR FOR SPIN AND CANONICAL energy-momentum current}\label{ETCR}
\subsection{Schwinger quantum action principle}
The second important tool required to construct the Sugawara-Sommerfield ansatz for the canonical currents of spin are their ETCRs. \par
We start with the transition amplitude which is equivalent to the sum over all possible histories of fields from the instant $\sigma_1$ to the instant $\sigma_2$  \cite{Schweber7783},
\begin{equation*}
\bra{\Psi_2;\sigma_2}\ket{\Psi_1;\sigma_1}  = N\sum_{H}\exp{\frac{i}{\hbar}I_H(\Omega)}.
\end{equation*}
Where $H$ corresponds to the history of evolution of an initial state (field) $\Psi_1$(x) corresponding to some value on $\sigma_1$ to a state $\Psi_2$(x) taking values on $\sigma_2$ and the other values between them $N$ is the normalization factor assuring, 
\begin{equation*}
\sum|\bra{\Psi_2;\sigma_2}\ket{\Psi_1;\sigma_1}|^2 = 1.    
\end{equation*} 
The \textit{Schwinger quantum action principle} states that an infinitesimal variation of the transition amplitude is equivalent to the matrix element of the variation of action by the infinitesimal parameters \cite{Schweber7783},
\begin{equation}\label{paper01_eq:12}
\delta\bra{\Psi_2;\sigma_2}\ket{\Psi_1;\sigma_1}=\bra{\Psi_2;\sigma_2}\frac{i}{\hbar}\int_{\sigma_1}^{\sigma_2} \delta\mathfrak{L}(\Psi,D\Psi) d^4x \ket{\Psi_1;\sigma_1} .
\end{equation} 
The ETCRs for the spin current can be derived using the Schwinger quantum action principle. Performing a delta variation w.r.t.\ the affine spin (Lorentz) connection on the matrix elements of the second Noether identity - Eq.(\ref{paper01_eq:11}), the canonical spin current Lie algebra then reads,\footnote{The spin current Lie-algebra corresponds to $\mathfrak{so}(1,3)\otimes \mathfrak{so}(1,3)$.}
\begin{align}\label{paper01_eq:17}
[\tensor{\mathfrak S}{_\alpha_\beta^0}(x),\tensor{\mathfrak S}{_\gamma_\delta^j}(x')]_{x_0=x'_0} &= 2i \big(\tensor{\eta}{_\gamma_[_\alpha}\tensor{\mathfrak S}{_\beta_]_\delta^j}(x)-\tensor{\eta}{_\delta_[_\alpha}\tensor{\mathfrak S}{_\beta_]_\gamma^j}(x)\big)\delta^3(x-x')\nonumber \\ 
&+i\bigg(\partial_i\frac{\delta\tensor{\mathfrak {S}}{_\alpha_\beta^i}(x)}{\delta\tensor{\Gamma}{_j^\gamma^\delta}(x')}-2\frac{\delta\mathfrak 
{T}_{[\alpha\beta]}(x)}{\delta\tensor{\Gamma}{_j^\gamma^\delta}(x')}\bigg).
\end{align}
 The second bracket in Eq. (\ref{paper01_eq:17}) represents the model dependent Schwinger terms.
\cleardoublepage

\subsection{Canonical energy-momentum current commutators}
The set of local ETCRs for the canonical energy-momentum current can be obtained using the same machinery by varying the matrix elements of the first Noether identity - Eq.(\ref{paper01_eq:10}) w.r.t.\ the coframe field.
Putting all this together we get the ETCR for the canonical energy-momentum current,
\begin{multline}\label{paper01_eq:20}
i[\tensor{\mathfrak{T}}{_\alpha^0}(x),\tensor{\mathfrak{T}}{_\theta^j}(x')]_{x_0=x'_0} =  \big(2\tensor{\mathfrak{T}}{_\theta^i}(x)\tensor*{\delta}{_[_\alpha^j}\tensor{\partial}{_i_]} + \frac{1}{2}\tensor{R}{_\alpha_\theta^\beta^\gamma}(x)\tensor{\mathfrak{S}}{_\beta_\gamma^j}(x)\big)\delta^3(x-x') \\
-\Bigg(\big(D_i-\tensor{T}{_\alpha_i^\beta}(x)\big)\frac{\delta\tensor{\mathfrak{T}}{_\beta^i}(x)}{\delta \tensor{e}{_j^\theta}(x')}
 - \frac{1}{2} \tensor{R}{_\alpha_i^\beta^\gamma}(x)\frac{\delta \tensor{\mathfrak{S}}{_\beta_\gamma^i}(x)}{\delta\tensor{e}{^\theta_j}(x')}\Bigg). \end{multline}
The terms in the big bracket of r.h.s. of Eq.\ (\ref{paper01_eq:20}) are the model dependent Schwinger terms. In the limit $\tensor{e}{^i_\alpha}=\tensor*{\delta}{^i_\alpha}$, $\tensor{\Gamma}{_i}=0$ (Equivalence principle for EC theory), it follows that the time-time components (energy density) commutators close on the momentum density,
\begin{equation}\label{paper01_eq:21}
[\tensor{\mathfrak{T}}{_0^0}(x),\tensor{\mathfrak{T}}{_0^0}(x')]_{x_0=x'_0}=-i\bigg(\tensor{\mathfrak{T}}{_0^a}(x)+\tensor{\mathfrak{T}}{_0^a}(x')\bigg)\partial_a\delta^3(x-x') - \tensor{\mathfrak{\Lambda}}{_{00}^{00}}(x,x').
\end{equation}
The Eq.(\ref{paper01_eq:21}) is analogous to Eq.(\ref{paper01_eq:7}a). Although it is important to note that in the EC theory the canonical energy-momentum encapsulates, apart from the distribution energy-matter distribution, an antisymmetric piece which is contributed by the spin of the particle. Hence, a correct formulation of the energy-momentum tensor for the strong and weak interactions describing particles carrying an intrinsic angular momentum should be in terms of the \textit{asymmetric} $\mathfrak{T}(x)$. Since the canonical energy-momentum current is the current coupled to the translation group $T_4$ , we call it the $\mathfrak{t}_4 \otimes \mathfrak{t}_4$ algebra. Note that the group of translations is not compact, although we use a similar notation for its algebra like current algebra.
\cleardoublepage
\section{ANALOGY BETWEEN SPIN CURRENT ALGEBRA AND VECTOR - AXIAL VECTOR CURRENT ALGEBRA} \label{comparison CSA}
The spin current Lie-algebra is very similar to the algebra of representation of the Lorentz group Eq.\ (\ref{paper01_eq:8}c), as it is expected. The role of the $SU(2)$  matrices $T^A$ in the current algebra is played by $SO(1,3)$ generators $\Sigma_{\alpha\beta}$. The Schwinger terms integrated should vanish, but in general are non-zero. For the sake of brevity, we record the different components of Eq.\  (\ref{paper01_eq:17}) in order to highlight the similarity with the $\mathfrak{su}(2)$ - currents.
\newline
The time-time components exhibit a similar algebra (ETCRs) to the charge density algebra -  Eq.\  (\ref{paper01_eq:1}),
\begin{subequations}\label{paper01_eq:22}
\begin{equation}
[\tensor{\mathfrak S}{_a_b^0}(x),\tensor{\mathfrak S}{_c_d^0}(x')]=2i\big(\tensor{\eta}{_c_[_a}\tensor{\mathfrak S}{_b_]_d^0}(x)-\tensor{\eta}{_d_[_a}\tensor{\mathfrak S}{_b_]_c^0}(x)\big)\delta^3(x-x')-\tensor{W}{_{ab,cd}^{0,0}}(x,x').
\end{equation}   
\begin{equation}
[\tensor{\mathfrak S}{_a_b^0}(x),\tensor{\mathfrak S}{_c_0^0}(x')]= 2i \tensor{\eta}{_c_[_a}\tensor{\mathfrak{S}}{_b_]_0^0} \delta^3(x-x')-\tensor{W}{_{ab,c0}^{0,0}}(x,x'),  
\end{equation}
\begin{equation}
[\tensor{\mathfrak{S}}{_a_0^0}(x),\tensor{\mathfrak {S}}{_b_0^0}(x')] = i\tensor{\mathfrak{ S}}{_a_b^0}(x)\delta^3(x-x') - \tensor{W}{_{a0,b0}^{0,0}}(x,x').
\end{equation}
\end{subequations}
Similarly, the time-space commutators are of a similar form as in Eq.\  (\ref{paper01_eq:2}),
\begin{subequations}\label{paper01_eq:23}
\begin{equation}
[\tensor{\mathfrak S}{_a_b^0}(x),\tensor{\mathfrak S}{_c_d^e}(x')]=2i\big(\tensor{\eta}{_c_[_a}\tensor{\mathfrak S}{_b_]_d^e}(x)-\tensor{\eta}{_d_[_a}\tensor{\mathfrak S}{_b_]_c^e}(x)\big)\delta^3(x-x')-\tensor{W}{_{ab,cd}^{0,e}}(x,x'),
\end{equation}
\begin{equation}
[\tensor{\mathfrak{S}}{_a_0^0}(x),\tensor{\mathfrak {S}}{_b_0^e}(x')] = i\tensor{\mathfrak{ S}}{_a_b^e}(x)\delta^3(x-x') - \tensor{W}{_{a0,b0}^{0,e}}(x,x'), \end{equation}
\begin{equation}
[\tensor{\mathfrak S}{_a_b^0}(x),\tensor{\mathfrak S}{_c_0^e}(x')]= 2i \tensor{\eta}{_c_[_a}\tensor{\mathfrak{S}}{_b_]_0^e} \delta^3(x-x')-\tensor{W}{_{ab,c0}^{0,e}}(x,x'),
\end{equation}
\begin{equation}
[\tensor{\mathfrak S}{_a_0^0}(x),\tensor{\mathfrak S}{_b_c^e}(x')]= 2i \tensor{\eta}{_a_[_b}\tensor{\mathfrak{S}}{_c_]_0^e} \delta^3(x-x')-\tensor{W}{_{ab,c0}^{0,e}}(x,x').
\end{equation}
\end{subequations}
Here $W$ are the model dependent Schwinger terms.
The space-space components identically vanish like in this case too.

Since the Sugawara-Sommerfield construction is built over the algebra of fields, the currents can be identified with its corresponding gauge fields  \cite{PhysRevLett.18.1029}. 
This allows us to express the canonical spin current (dynamical variables for gravity) in terms of the Lorentz connection,\footnote{The model dependent terms $W$ are obtained by substituting Eq.\  (\ref{CP}) into the spin current algebra.}
\begin{equation}\label{CP}
\tensor{\mathfrak{S}}{_\alpha_\beta_i}(x) :=  C \tensor {\Gamma}{_i_\alpha_\beta}(x), 
\end{equation}
where $C = \frac{m^2}{g^2}$. Hence, the spin current algebra pertaining to the Sugawara-Sommerfield model is
\begin{align}\label{new}
[\tensor{\mathfrak S}{_\alpha_\beta_0}(x),\tensor{\mathfrak S}{_\gamma_\delta_j}(x')] &=2i \big(\tensor{\eta}{_\gamma_[_\alpha}\tensor{\mathfrak S}{_\beta_]_\delta_j}(x)-\tensor{\eta}{_\delta_[_\alpha}\tensor{\mathfrak S}{_\beta_]_\gamma_j}(x)\big)\delta^3(x-x')\nonumber \\ 
&+ 2iC\eta_{\alpha[\gamma}\eta_{\delta]\beta}\partial_j\delta^3(x-x'\big).
\end{align}
The coefficients of the Schwinger terms are the $SO(1,3)$ Cartan-metric $2\eta_{\alpha[\gamma}\eta_{\delta]\beta}$ (compare Eqs.\ (\ref{paper01_eq:1}, \ref{paper01_eq:2}) having the $SU(2)$ metric $\tensor{\delta}{^A^B}$ as coefficients of the Schwinger terms.)
\subsection{The analogy between the Lorentz algebra and vector-axial vector currents}
The canonical spin tensor corresponding to the rotations (spatial) Lorentz indices $\tensor{\mathfrak{S}}{_a_b^i}$ will be denoted as $\mathfrak{S}_r$. Similarly, $\tensor{\mathfrak{S}}{_c_0^i}$ corresponding to  boosts (temporal-spatial) components will be denoted as $\mathfrak{S}_b$.
Comparing the local equal-time commutation relations for the spin tensor - Eqs.\ (\ref{paper01_eq:22}, \ref{paper01_eq:23}) and the hadronic currents - Eqs.\ (\ref{paper01_eq:1}, \ref{paper01_eq:2}) we find that,
\begin{align}\label{comparison}
[\mathfrak{S}_r, \mathfrak{S}_r] \propto \mathfrak{S}_r &&  [J, J] \propto J ,\\ \nonumber
[\mathfrak{S}_r, \mathfrak{S}_b] \propto \mathfrak{S}_b && [J, J_5] \propto J_5,\\ \nonumber
[\mathfrak{S}_b, \mathfrak{S}_b] \propto \mathfrak{S}_r && [J_5, J_5] \propto J, 
\end{align}
\textit{The closure of the algebras indicate that  $\mathfrak{S}_r \propto J$ alone while $\mathfrak{S}_b \propto J_5$.} Thus the canonical Noether current associated with Lorentz symmetry are interlocked with Noether current associated with internal symmetry.
\section{SUGAWARA-SOMMERFIELD CONSTRUCTION FOR spin current}
In order to achieve this, the spin current ansatz should satisfy the following conditions: \footnote{Such a construction is equally applicable for $SU(3)_f$ group.}

\textbf{(i)} It should be a \textit{unitary singlet} under the $SU(2)_f$ transformation.

\textbf{(ii)} The ansatz should satisfy the spin current Lie-Algebra - Eq.\  (\ref{new}) and the hadronic current algebra.
\newline

The \textit{Belinfante-Rosenfeld} formula in the framework of EC theory expresses the auxiliary symmetric energy-momentum tensor in terms of the canonical energy-momentum tensor and the gradients of spin tensor,
 \newline 
 \begin{equation}\label{BR formula}
\tensor{\Theta}{_\alpha^i} (x) = \tensor{\mathfrak{T}}{_\alpha^i} (x) - \frac{1}{2} \overset{*}{D_k}\big(\tensor{\mathfrak{S}}{_\alpha^i^k}(x)-\tensor{\mathfrak{S}}{^i^k_\alpha}(x)+\tensor{\mathfrak{S}}{^i_\alpha^k}(x)\big) .
\end{equation}
\newline 
Here, $\overset{*}{D_k} = D_k + \tensor{T}{_k_l^l}$ is the modified covariant derivative. Considering only the partial derivative part we infer from Eqs.\ (\ref{paper01_eq:5a}, \ref{paper01_eq:5b}) and Eq. (\ref{BR formula}) that the components of the canonical spin current can be at most \textit{linear} in the hadronic currents (cf. Appendix \ref{Appendix A} for a current algebra argument).
\newline 

Our inference from Eqs.(\ref{comparison}) that   $\mathfrak{S}_r$ is solely related to $J$ and $\mathfrak{S}_b$ to $J_5$, a linear polynomial in hadronic currents and the singlet condition put together yields separate equations for the rotation and boost components of the spin current,
\begin{subequations}\label{s2c}
\begin{align}
\tensor{\mathfrak{S}}{_a_b_i}(x) &= \tensor{\chi}{_a_b^A}(x)\tensor{J}{^A_i}(x) = \Tr_{SU(2)}\big(\tensor{\chi}{_a_b}(x)\tensor{J}{_i}(x)\big).
\\
\tensor{\mathfrak{S}}{_c_0_i}(x) &= \tensor{\chi}{_c_0^A}(x)\tensor{J}{_5^A_i}(x) = \Tr_{SU(2)}\big(\tensor{\chi}{_c_0}(x)\tensor{J}{_5_i}(x)\big).\end{align}
\end{subequations}
Where $\tensor{\chi}{_i_j^A}(x) = -\tensor{\chi}{_j_i^A}(x)$ are components of a \textit{skew-symmetric} $2$-form. It is straight forward to work with Eq.(\ref{s2c}a) and fix the form of the spatial parts, namely, $\tensor{\chi}{_a_b^A}$.\footnote{The form of $\chi_{c0}$ in Eq. (\ref{s2c}b) is complicated due to being related to $J_{5i}$, which, is related to vierbiens corresponding to the identity generator of the extended $U(2)$ flavour group (cf. \cite{PhysRev.172.1542}).} It is obtained by substituting Eq. (\ref{s2c}a) into Eq. (\ref{new}) in order to exactly reproduce its right hand side. Thus, $\chi$ takes the form, 
\newline
\begin{equation} \label{Main}
\tensor{\chi}{_a_b^A}(x)=\frac{1}{2}\tensor{\epsilon}{_A_B_C} \big(\tensor{E}{_a^B}(x)\tensor{E}{_b^C}(x)-\tensor{E}{_a^B}(x)\tensor{E}{_b^C}(x)\big) = \tensor{\epsilon}{_A_B_C} \tensor{E}{_[_a^B}\tensor{E}{_b_]^C}(x).\end{equation}
\newline 
Here we define $\mathfrak{su}(2)$-valued $1$-forms (coframe with anholonomic flavour indices),
\begin{equation}\label{intvier}
\vartheta^A(x) = \tensor{E}{_i^A}(x) dx^i.   
\end{equation}
The components $\tensor{E}{_i^A}$(x) could be called \enquote{internal-vierbeins}. Such vierbeins appear in \citep{PhysRev.172.1542, article}. It is also important to note that these coframes are related to $SU(2)_f$ (flavour) and not to $SO(1,3)$ freedom like the convential coframes. Also, one \textit{cannot} convert from space-time to internal indices using these quantities.
\begin{equation}
\vartheta_i(x)=\tensor{E}{_i^A}(x)T_A.
\end{equation}
These vierbeins are orthogonal matrices that satisfy the following contraction properties,%\footnote{For $SU(3)$ the contraction is $\tensor{e}{_i^A}(x)\tensor{e}{_j^A}(x)=\eta_{ij}$, $\tensor{e}{_i^A}(x)\tensor{e}{^i^B}(x)=\sqrt{\frac{3}{2}}\delta^{AB}$}
\begin{align*}
\tensor{E}{_i^A}(x)\tensor{E}{_j^A}(x)=\eta_{ij}, &&  \tensor{E}{_i^A}(x)\tensor{E}{^i^B}(x)=\delta^{AB}.
\end{align*}
Thus, the \textit{Sugawara-Sommerfield constuction for the rotation components of the canonical spin current} reads
\begin{equation}\label{main}
\tensor{\mathfrak{S}}{_a_b_i}(x) = \tensor{\epsilon}{_A_B_C} \tensor{E}{_[_a^A}(x)\tensor{E}{_b_]^B}(x) \tensor{J}{^C_i}(x).
\end{equation}
%Using the analogy set up in Eq.\  (\ref{comparison}), we can split the rotation and the boost parts of the spin current,
%\begin{subequations}\label{RVBC}
%\begin{align}
%\tensor{\mathfrak{S}}{_a_b_i}(x)=\frac{1}{2}\tensor{\epsilon}{_A_B_C} \big(\tensor{E}{_a^A}(x)\tensor{E}{_b^B}(x)-\tensor{E}{_b^A}(x)\tensor{E}{_a^B}(x)\big)\tensor{J}{^C_i}(x),\\
%\tensor{\mathfrak{S}}{_c_0_i}(x)=\frac{1}{2}\tensor{\epsilon}{_A_B_C} \big(\tensor{E}{_c^A}(x)\tensor{E}{_0^B}(x)-\tensor{E}{_0^A}(x)\tensor{E}{_c^B}(x)\big) \tensor{J}{_5^C_i}(x).
%\end{align}
%\end{subequations}
\subsection{Pure gauge form}
The vector currents of the Sugawara-Sommerfield model have a solution in the form of \citep{PhysRev.172.1542, Tauber1974}
\begin{equation}\label{currep}
\tensor{J}{^A_i}(x) = \frac{1}{2}C\tensor{\epsilon}{^A^B^C}\tensor{E}{^B_l}(x)\partial_i\tensor{E}{^C^l}(x).
\end{equation}
Substituting Eq. (\ref{currep}) into Eq. (\ref{main}), we get a singlet in the internal-dreibeins, 
\begin{equation}\label{spinvier}
\tensor{\mathfrak{S}}{_a_b^i}(x)= C \tensor{E}{_A_[_a}\partial^i\tensor{E}{_b_]^A}(x). 
\end{equation}
This is of the form $E^{-1}d E$. Hence, the spin tensor assumes a pure gauge form in this model. It follows that the four-divergence of the spin current is not zero but rather yields the \textit{antisymmetric} part of the canonical energy-momentum current,
\begin{equation}\label{eq:noncon}
\partial_i\tensor{\mathfrak{S}}{_a_b^i}(x)= C \tensor{E}{_A_[_a}\Box\tensor{E}{_b_]^A} = 2\tensor{\mathfrak{T}}{_[_a_b_]}.
\end{equation}
Here $\Box = \partial_i\partial^i$ is the D’Alembert operator.

\subsection{Internal dreibein quantization}
Due to the pure gauge form it is easy to see that the internal-vierbiens satisfy a set of canonical commutation relations.  Substituting Eq.\ (\ref{spinvier}) into Eq.\ (\ref{new}) leads to the following \textit{canonical} commutation rules à la Yang-Mills-Ashtekar for the internal dreibeins: 
\begin{subequations}\label{vierquant}
\begin{equation}
[\tensor{E}{_a^A}(x), \tensor{E}{_b^B}(x')]_{x_0=x'_0} = 0, 
\end{equation}
\begin{equation}
[\tensor{E}{_a^A}(x),\tensor{\Pi}{_0_b^B}(x')]_{x_0=x'_0} = -\frac{i}{C}\tensor{\delta}{^A^B}\tensor{\eta}{_a_b}\delta^3(x-x'),    
\end{equation}
\begin{equation}
[\tensor{\Pi}{_0_a^A}(x),\tensor{\Pi}{_0_b^B}(x')]_{x_0=x'_0} = 0,    
\end{equation}
\begin{equation}
 [\tensor{\Pi}{_0_a^A}(x),\tensor{\Pi}{_b_c^B}(x')]_{x_0=x'_0} =  \frac{2i}{C}\delta^{AB}\eta_{a[b}\partial_{c]}\delta^3(x-x'). \end{equation}
\end{subequations}
Where $\tensor{\Pi}{_0_a^A}(x) = \partial_0\tensor{E}{_a^A}(x)$  are the \textit{canonically conjugate momenta} and \newline 
$\tensor{\Pi}{_a_b^A}(x) = \partial_a\tensor{E}{_b^A}(x)$ . This leads to a consistent spin current Lie-algebra (see Appendix \ref{Appendix C} for calculations). 
\subsection{Gauge potentials interlink}

Using Eq.\ (\ref{CP}), the $SO(3)$ part (rotation Lorentz indices) of the spin connection can be solely expressed in terms of the $SU(2)$ (isospin) Yang-Mills gauge connection,
\begin{equation}\label{Lo2ga}
\tensor{\Gamma}{_i_a_b}(x) =\tensor{\epsilon}{_A_B_C}\tensor{E}{_[_a^A}(x)\tensor{E}{_b_]^B}(x)\tensor{A}{_i^C}(x).
\end{equation}
Thus the \textit{gravitational gauge potential} $\Gamma(x)$ \textit{is built out of the non-Abelian Yang-Mills gauge potential} $A(x)$. 
\newline
The curvature of the affine spin connection - Eq.\  (\ref{Lo2ga}), vanishes identically. Similar results were obtained in \cite{10.1093/ptep/ptaa108}. (see, also Appendix \ref{Appendix D} for calculations)\footnote{Even though cumbersome, it should be possible to show that the Lorentz curvature with temporal-space indices also vanish, hence leading to a completely vanishing curvature term.},
\begin{equation}
\tensor{R}{_a_b_i_j} (x)=0.
\end{equation}
This is an interesting result since we find a vanishing Lorentz curvature tensor and a non-vanishing torsion tensor. Therefore, in the $m=0$, $g=0$ limit we find the emergence of a \textit{teleparallelism}\footnote{Einstein had tried setting up a unified theory of gravity and electromagnetism based on \enquote {Fernparallelismus/ teleparallelism}.} in the local geometry of strong interactions. Hence, as we approach this \enquote{formal} limit, the current algebra leads to the evolution of local geometry from a Riemann-Cartan (RC) spacetime to a \textit{Weitzenböck spacetime} and the connection is a Weitzenböck connection. 
\newline

Suprisingly, similar feature arises while studying two dimensional systems, namely the non-linear $\sigma$ model with WZW interactions. In this case torsion is naturally incorporated to the geometry along with curvature. On a renormalization flow of the geometry of the manifold, there arises non-trivial infrared fixed points $\lambda = \pm 4\pi /N$ for $N \in \mathbb{Z}$. At these special values of the coupling $\lambda$, the manifold is \enquote{parallelizable} and the geometry evolves from RC to Weitzenböck. This concept was first introduced in \citep{BRAATEN1985630, PhysRevLett.53.1799} and is called \enquote{Geometrostasis}. It would be very interesting to see the connection between the SS construction for spin current in 4-dimensions and the 2-dimensional model studied in \cite{BRAATEN1985630}.

\subsection{An extended ansatz}
%Does the spin current ansatz - Eq.\  (\ref{spinvier}) admit other terms to be added such that it is consistent with the spin current Lie-algebra?
Apart from Eq.\ (\ref{vierquant}), we calculate the ETCRs for the time-time and time-space components.
\begin{subequations}\label{t-s vier}
\begin{equation}
[\tensor{E}{_0^A}(x), \tensor{E}{_0^B}(x')]_{x_0=x'_0} = \frac{i}{2\kappa C^{\frac{3}{2}}}\tensor{\epsilon}{_A_B_C}\tensor{E}{_0^C}(x)\delta^3(x-x'),\end{equation}
\begin{equation}
[\tensor{E}{_0^A}(x), \tensor{E}{_a^B}(x')]_{x_0=x'_0} = \frac{i}{2\kappa C^{\frac{3}{2}}}\tensor{\epsilon}{_A_B_C}\tensor{E}{_a^C}(x)\delta^3(x-x') + \frac{i\tensor{\delta}{^A^B}}{2\kappa C^2}\partial_a\delta^3(x-x').      
\end{equation}
\end{subequations}
These commutators are similar to Eqs.\ (\ref{paper01_eq:1}, \ref{paper01_eq:2}) apart from a factor of $C^{\frac{3}{2}}$ in the denominator which ensures correct length dimensions. Even though, an explicit proof is missing, one expects these commutators to hold since it is based on current algebra again.\footnote{It would be important to see if the above commutators can be derived from the Lagrangian formulation.}
\newline
With these commutation rules, it is easy to show that the spin tensor admits a unique \textit{cubic} term in $\tensor{E}{_a^A}(x)$ and the complete ansatz reads,
%\begin{equation}
%\partial_i\tensor{e}{_a^A}(x) = \frac{2}{3}\tensor{\epsilon}{_A_B_C} \tensor{e}{_a^A}(x)\tensor{e}{_a^A}(x)\tensor{e}{_a^A}(x)   
%\end{equation}
%\textit{unique} term to Eq.\  (\ref{spinvier}); so as to recover.  One finds that it admits a term precisely  Thus, the complete spin current solution reads,\footnote{This Chern-Simons like $3$-form term can be obtained after some calculation using the commutation rules - Eqs.\ (\ref{vierquant}, \ref{t-s vier}) yield us the cubic extension.}
\begin{equation}\label{CSSf}
\tensor{\mathfrak{S}}{_a_b^i}(x)= \tensor{\widehat{E}}{_A_[_a}\partial^i\tensor{\widehat{E}}{_b_]^A}(x) + \frac{2}{3} \tensor{\epsilon}{_A_B_C}\tensor{\widehat{E}}{_[_a^A}(x)\tensor{\widehat{E}}{_b_]^B}(x)\tensor{\widehat{E}}{^i^C}(x). \end{equation}

Where $\tensor{\widehat{E}}{_i^A}(x) := C^{\frac{1}{2}}\tensor{E}{_i^A}(x)$ are a rescaled set of vierbeins and $C^{\frac{1}{2}}$ has an inverse length dimension,i.e., $d_{C^{\frac{1}{2}}} = -1$.  

\section{DISCUSSION AND CONCLUSION}
We showed that the canonical spin current (gravitational currents associated to Lorentz symmetry) can be constructed out of the hadronic flavour currents (internal symmetry). Rather, one learns by comparing the $SO(1,3)$ current algebra and the $SU(2)_f$ current/charge algebra that the Lorentz Boosts can be interlinked to the flavour chiral (axial) charges  $\Sigma_{0a} := K_a \propto Q_5$ and $\mathfrak{S}_b$ are related to \textit{axial vector currents}. On the other hand, Lorentz rotations are interlinked to the flavour charges $S_a := \frac{1}{2} \tensor{\epsilon}{_a_b_c}\Sigma_{bc} \propto Q$ and $\mathfrak{S}_r$ are related to \textit{vector currents}. 
%\textbf{(ii)} The ansatz we have constructed resembles the \textit{Mathisson-Weysenhoff} ansatz describing the spin tensor of classical massive spinning fluids (cf. \cite{Obukhov_1993}).
\par
As a consequence of the currents being scaled gauge fields, the spin construction in terms of the currents has surprisingly led to an equation interlinking the \textit{spin (Lorentz) connection and the Yang-Mills gauge connection}.

%\textbf{(iv)} The linear dependence of spin on currents - Eq.\  (\ref{main}) leads to an  introduction of $\mathfrak{su}(2)$-valued vierbeins. Their antisymmetrized product has units spin/charge. If the currents are constructed like - Eq.\  (\ref{currep}), it follows that $\tensor{e}{_i^A}(x)$ obey a \textit{Yang-Mills-Ashtekar} like quantization.\footnote{I have also added the name Ashtekar for this quantization process. Although, the quantization of the triads in Ashtekar's theory is w.r.t.\ the $SO(3)$ external Lorentz symmetry unlike our case.}
An interesting result is the pure gauge form of the spin current/connection $\tensor{\Gamma}{_i_a_b}(x) =\tensor{E}{_A_[_a}\partial_i\tensor{E}{_b_]^A}(x)$ (Weitzenböck type) which has a vanishing Lorentz curvature but non-vanishing torsion tensor (Teleparallelism). This changes the local geometry from a Riemann-Cartan spacetime to a \textit{Weitzenböck} spacetime. If we switch from a 4-dimensional to a 2-dimensional case, our construction seems to be related to the works in \cite{BRAATEN1985630}, where teleparellism occurs at special values of the couplings $\lambda = \pm 4\pi /N$ for $N \in \mathbb{Z}$. These are the non-trivial infrared fixed points of their model. A further investigation is needed to verify the equivalence of this extended SS construction to the non-linear $\sigma$ model with WZW interactions.

It follows from the spin current ansatz that the internal-vierbeins satisty a set of canonical commutation rules à la Yang-Mills - Eqs. (\ref{vierquant}), indicating that the Sugawara-Sommerfield model is indeed a canonical formalism in terms of these internal-vierbeins.
%\textbf{(vi)} There is another distinct difference between the classical and canonical approach. Both approaches lead to a vanishing curvature tensor. In the standard approach choosing a frame $e=\delta$ and the current equations of motion lead to a vanishing curvature (see Appendix \ref{Appendix D}). Also, one gets a spin current\textit{
%conservation} in this approach. This is completely circumvented by the canonical approach in which $e(x)$ fields and its conjugate momenta $\Pi(x)$ satisfy canonical commutation rules; automatically yielding a vanishing curvature. In contrast to the previous approach, the canonical approach leads to a non-vanishing spin current divergence - Eq.\  (\ref{eq:noncon}) and is equal to the antisymmetric part of the energy-momentum currents.\footnote{In this manuscript I have managed only for the spatial part of the Lorentz indices. An extension to temporal components would be done in future} 
Lastly, we realise that the vanishing Lorentz curvature, $\tensor{R}{_a_b^i^j}(x) = 0$ is a consequence of a vanishing Yang-Mills field strength $\lim_{m\rightarrow 0, g\rightarrow 0} \tensor{\widetilde{F}}{_i_j^A}(x) = 0$. This means that the limit $m,g \rightarrow 0$ has a major role in the manifold being \enquote{parallelized}.
\par 
\subsection{Some potential applications of our construction}
Even though, the original SS construction had not found direct experimental evidences in the high-energy physics regime, it has found some very useful applications in 2-d Conformal Field Theories (CFT). Since, we are dealing with spin-charge-flavour currents, it is indicative that one can apply our construction for spin currents in CFT (for e.g. the \enquote{spin quantum Hall effect} or the \enquote{Kondo model}).
\par
Also, there has been a recent interesting work \cite{2021}, where the Abelian $U(1)$ electromagnetic vector potential $A_i(x)$ serves as the spin connection for the \textit{Bogoliubov fermionic quasiparticles in the chiral superconductors}. This could be seen as an Abelian counterpart of our work, namely, the rotational gauge potential-Yang-Mills gauge potential identification (although, this has nothing to do with the SS construction). This motivates us to also look into condensed matter systems for further experimental investigations.  
 
\begin{acknowledgments}
I am indebted to my supervisor Prof.\ Dr.\ Claus Kiefer who constantly guided and supported me to work on this model accompanied by many helpful discussions. I am also indebted to Prof.\ Dr.\ Friedrich W. Hehl for not only bringing this project underway but also for consistently guiding me with numerous insightful discussions and helpful remarks. Both of them also motivated me to publish my results. This project was partly supported by the Bonn–Cologne Graduate School of Physics and Astronomy scholarship.
\end{acknowledgments}
\appendix

\section{Spin tensor as a linear polynomial}\label{Appendix A}
We begin with the Lagrangian density for a massive Yang-Mills theory (spin-1 massive gauge fields) in a RC spacetime,
\begin{equation}
\mathfrak{L}(A,\partial A, \Gamma ,x)=-\frac{1}{4}\tensor{F}{_i_j^A}(x)\tensor{F}{^i^j^A}(x)+\frac{m^2}{2}\tensor{A}{_i^A}(x)\tensor{A}{^i^A}(x),
\end{equation}
where,
\begin{align}\label{fsrc}
\tensor{F}{_i_j^A}(x)& := 2\widetilde{D}_{[i}\tensor{A}{_j_]^A} &=  2\partial_{[i} \tensor{A}{_j_]^A}(x) -\frac{1}{2}g\tensor{\epsilon}{_A_B_C}\{\tensor{A}{_i^B}(x),\tensor{A}{_j^C}(x)\}+2\tensor{\Gamma}{_[_i^u^v}(x)\big(\tensor{\Sigma}{_u_v})_{j]}^p\tensor{A}{_p^A}(x).
\end{align}
Where $(\tensor{\Sigma}{_u_v})_j{}^p$ is the vector representation of the Lorentz group.
Consider the ETCRs for the gauge fields \cite{PhysRevLett.18.1029}. For simplicity we  avoid the contribution of the last term in Eq. (\ref{fsrc}),
\begin{subequations}\label{oldcomm}
\begin{equation}
[\tensor{A}{_0^A}(x),\tensor{A}{_0^B}(x')]= \frac{ig}{m^2}\tensor{\epsilon}{_A_B_C}\tensor{A}{_0^C}(x)\delta^3(x-x'),
\end{equation}
\begin{equation}
[\tensor{A}{_0^A}(x),\tensor{A}{_i^B}(x')]= \frac{ig}{m^2}\tensor{\epsilon}{_A_B_C } \tensor{A}{_i^C}(x)\delta^3(x-x')+\frac{i}{m^2}\tensor{\delta}{^A^B}\partial_i \delta^3(x-x'), 
\end{equation}
\begin{equation}
[\partial_0\tensor{A}{_i^A}(x)- \partial_i\tensor{A}{_0^A}(x),\tensor{A}{_j^B}(x')]= i\eta_{ij}\delta^{AB}\delta^3(x-x')+\frac{ig}{m^2}\tensor{\epsilon}{_A_B_C} \tensor{A}{_i^C}(x)\partial_j\delta^3(x-x')+\mathcal{O}(A^2)\delta^3(x-x').    
\end{equation}
\end{subequations}
The canonical spin tensor for a spin-1 massive gauge field is,
\begin{align}\label{gauge spin}
\tensor{\mathfrak{S}}{_i_j^k}(x) &= 2 \tensor{F}{^k_[_i^A}(x)\tensor{A}{_j_]^A}(x).
\end{align}
It is clear that the spin tensor contains a cubic term in the gauge fields. 
As prescribed in \cite{PhysRev.170.1353}, currents can be defined by scaling the gauge fields (operators),
\begin{equation*}
\tensor{J}{^A_i}(x)=\frac{m^2}{g}\tensor{A}{_i^A}(x)   
\end{equation*}
the corresponding scaled field strength is,
\begin{equation*}
\tensor{\widetilde{F}}{_i_j^A}(x)=\frac{m^2}{g}\tensor{F}{_i_j^A}(x).
\end{equation*}
The Sugawara model can be revived from the massive Yang-Mills theory by \cite{PhysRev.170.1353},
\begin{equation*}
\lim_{m\rightarrow 0, g\rightarrow 0,\frac{m^2}{g^2}=C} \tensor{\widetilde{F}}{_0_i^A}(x) = 0.    
\end{equation*}
The corresponding change in the algebra of fields is the current commutators Eq.(\ref{paper01_eq:2}),
\begin{subequations}\label{newcomm}
\begin{equation}
[\tensor{J}{^A_0}(x),\tensor{J}{^B_0}(x')]_{x_0=x'_0}= i\tensor{\epsilon}{_A_B_C } \tensor{J}{^C_0}(x)\delta^3(x-x'),
\end{equation}
\begin{equation}
[\tensor{J}{^A_0}(x),\tensor{J}{^B_i}(x')]_{x_0=x'_0}= i\tensor{\epsilon}{_A_B_C } \tensor{J}{^C_i}(x)\delta^3(x-x')+iC\tensor{\delta}{^A^B}\partial_i \delta^3(x-x'),    
\end{equation}
\begin{equation}
[\partial_0\tensor{J}{^A_i}(x)- \partial_i\tensor{J}{^A_0}(x),\tensor{J}{^B_j}(x')] = i\tensor{\epsilon}{_A_B_C} \tensor{J}{^C_i}(x)\partial_j\delta^3(x-x')+\mathcal{O}(J^2)\delta^3(x-x').    
\end{equation}
\end{subequations}

Where $\partial_0\tensor{J}{^A_i}(x)-\partial_i\tensor{J}{^A_0}(x)= \frac{1}{2C}\tensor{\epsilon}{_A_B_C}\{\tensor{J}{^B_0}(x),\tensor{J}{^C_i}(x)\}$. The difference between the quantization of Eq. (\ref{oldcomm}c) and Eq. (\ref{newcomm}c) is evident, since the latter is devoid of $i\eta_{ij}\delta^{AB}\delta^3(x-x')$. Let us assume that the spin current ansatz was a cubic polynomial like Eq. (\ref{gauge spin}) for e.g.,
\begin{equation*}
\tensor{\mathfrak{S}}{_0_i_k}(x)=\big(\partial_0\tensor{J}{^A_i}(x)-\partial_i\tensor{J}{^A_0}(x)\big)\tensor{J}{^A_k}(x)= \frac{1}{2C}\tensor{\epsilon}{_A_B_C}\{\tensor{J}{^B_0}(x),\tensor{J}{^C_i}(x)\}\tensor{J}{^A_k}(x).
\end{equation*} 
 On substituting the above expression into Eq. (\ref{paper01_eq:17}) and using  Eq. (\ref{newcomm}c), we either end up with fifth order terms as the coefficients of delta function or fourth-order terms as  coefficients of derivatives of delta function. This surely doesn't satisfy the spin current algebra. Hence, it can admit at most a \textit{linear} polynomial in the hadronic currents.
\cleardoublepage
\cleardoublepage
\section{Verifying the ansatz using internal-dreibein commutation rules}\label{Appendix C}
Firstly we show that the spin current completely expressed in terms of the $\mathfrak{su}(2)$ vierbeins satisfy the spin current Lie-algebra. Our calculations are only restricted to $\mathfrak{S}_r$.
\begin{equation*}
\tensor{\mathfrak{S}}{_a_b^0}(x) = C \tensor{E}{^A_[_a}\partial^0\tensor{E}{_b_]^A}(x),
\end{equation*}
\begin{equation*}
\tensor{\mathfrak{S}}{_c_d^e}(x') = C \tensor{E}{^B_[_c}\partial^e\tensor{E}{_d_]^B}(x').
\end{equation*}
Substituting in ETCR for spin we get,
\begin{align*}
[\tensor{\mathfrak{S}}{_a_b^0}(x),\tensor{\mathfrak{S}}{_c_d^e}(x')]_{x_0=x'_0} &= C^2[\tensor{E}{^A_[_a}\partial^0\tensor{E}{_b_]^A}(x),\tensor{E}{^B_[_c}\partial^e\tensor{E}{_d_]^B}(x')]\\
&= C^2[\tensor{E}{^A_a}\partial^0\tensor{E}{_b^A}(x)-\tensor{E}{^A_b}\partial^0\tensor{E}{_a^B}(x), \tensor{E}{^B_c}\partial^e\tensor{E}{_d^B}(x')-\tensor{E}{^B_d}\partial^e\tensor{E}{_c^B}(x')]\\
&=C^2\bigg([\tensor{E}{^A_a}\partial^0\tensor{E}{_b^A}(x), \tensor{E}{^B_c}\partial^e\tensor{E}{_d^B}(x')]- [\tensor{E}{^A_a}\partial^0\tensor{E}{_b^A}(x), \tensor{E}{^B_d}\partial^e\tensor{E}{_c^B}(x')]\\
&-[\tensor{E}{^A_b}\partial^0\tensor{E}{_a^B}(x), \tensor{E}{^B_c}\partial^e\tensor{E}{_d^B}(x')]+ [\tensor{E}{^A_b}\partial^0\tensor{E}{_a^B}(x),\tensor{E}{^B_d}\partial^e\tensor{E}{_c^B}(x')]\bigg).
\end{align*}
Using the commutation rules - Eq.  (\ref{vierquant}), we get our desired result,
\begin{align*}
&=i\big(\tensor{\eta}{_a_c}\tensor{\mathfrak S}{_b_d_e}(x)-\tensor{\eta}{_a_d}\tensor{\mathfrak S}{_b_c_e}(x)+\tensor{\eta}{_b_d}\tensor{\mathfrak S}{_a_c_e}(x)-\tensor{\eta}{_b_c}\tensor{\mathfrak S}{_a_d_e}(x)\big)\delta^3(x-x')\\
&+iC\big(\eta_{ac}\eta_{bd}-\eta_{ad}\eta_{bc})\partial_e\delta^3(x-x').    
\end{align*}
\cleardoublepage
\section{Vanishing Lorentz curvature}\label{Appendix D}
The Lorentz curvature is defined as,
\begin{equation*}
\tensor{R}{_i_j^a^b} := \partial_i\tensor{\Gamma}{_j^a^b} - \partial_j\tensor{\Gamma}{_i^a^b} + \tensor{\Gamma}{_i^a_c}\tensor{\Gamma}{_j^c^b}-\tensor{\Gamma}{_j^a_c} \tensor{\Gamma}{_i^c^b}
\end{equation*}

We start with Eq.  (\ref{Lo2ga}), $\tensor{\Gamma}{_i^a^b}(x) = 2 \tensor{E}{_A^[^a}\partial_i\tensor{E}{^b^]_A}(x)$.
\begin{align*}
\partial_i\tensor{\Gamma}{_j^a^b} - \partial_j\tensor{\Gamma}{_i^a^b} &= 2\partial_i\tensor{E}{_A^[^a}\partial_j\tensor{E}{^b^]_A} -   2\partial_j\tensor{E}{_A^[^a}\partial_i\tensor{E}{^b^]_A}.
\end{align*}
The product of the connections is,
\begin{align*}
\tensor{\Gamma}{_i^a_c}\tensor{\Gamma}{_j^c^b}-\tensor{\Gamma}{_j^a_c} \tensor{\Gamma}{_i^c^b} &= \eta_{cp}\big(\tensor{E}{_A^a}\partial_i\tensor{E}{^p_A}-\tensor{E}{_A^p}\partial_i\tensor{E}{^a_A})\big(\tensor{E}{_B^c}\partial_j\tensor{E}{^b_B}-\tensor{E}{_B^b}\partial_j\tensor{E}{^c_B}\big)\\
&- \eta_{cp}\big(\tensor{E}{_A^a}\partial_j\tensor{E}{_A^p}-\tensor{E}{^p_A}\partial_j\tensor{E}{^a_A})\big(\tensor{E}{_B^c}\partial_i\tensor{E}{^b_B}-\tensor{E}{^B^b}\partial_i\tensor{E}{^c_B}\big)\\
&=\tensor{E}{_A^a}\tensor{E}{_B_c} \partial_i\tensor{E}{_A^c}\partial_j\tensor{E}{^b_B}-\tensor{E}{_A^a}\tensor{E}{_B^b} \partial_i\tensor{E}{_A_c}\partial_j\tensor{E}{^c_B}-\partial_i\tensor{E}{^a_A}\partial_j\tensor{E}{^b_A}+\tensor{E}{_c_A}\tensor{E}{^b_B} \partial_i\tensor{E}{^a_A}\partial_j\tensor{E}{^c_B}\\
&-\tensor{E}{_A^a}\tensor{E}{_B^c} \partial_j\tensor{E}{_c_A}\partial_i\tensor{E}{^b_B}+\tensor{E}{_A^a}\tensor{E}{_B^b}\partial_j\tensor{E}{_c_A}\partial_i\tensor{E}{^b_B}+\partial_j\tensor{E}{^a_A}\partial_i\tensor{E}{^b_A}-\tensor{E}{_A_c}\tensor{E}{_B^b} \partial_j\tensor{E}{^a_A}\partial_i\tensor{E}{^c_B}\\
&= - 2\partial_i\tensor{E}{_A^[^a}\partial_j\tensor{E}{^b^]_A} + 2\partial_j\tensor{E}{_A^[^a}\partial_i\tensor{E}{^b^]_A}.
\end{align*}
There is a nice cancellation among the terms due to the skew-symmetry of the spin (Lorentz) connection.
Hence we get $\tensor{R}{_a_b_i_j}=0$ (vanishing Lorentz curvature).
\bibliography{Bib}% Produces the bibliography via BibTeX.
\end{document}